\documentclass[letterpaper,twocolumn,prl,preprintnumbers,showpacs,nofootinbib]{revtex4}
\usepackage{graphicx, amsmath, amssymb, bm, epstopdf,float, relsize, mathrsfs, dsfont, amsfonts, scalefnt}
\DeclareMathSizes{7}{5}{4}{3} 
\DeclareMathSizes{5}{3}{3}{2} 

\begin{document}

\title{Intrinsic upper bound on two-qubit polarization entanglement predetermined by pump polarization correlations in
parametric down-conversion}

\author{Girish Kulkarni, V. Subrahmanyam, and Anand K. Jha}

\email{akjha9@gmail.com}

\affiliation{Department of Physics, Indian Institute of Technology, Kanpur 208016, India}

\date{\today}

\begin{abstract}

We study how one-particle correlations transfer to manifest as
two-particle correlations in the context of parametric
down-conversion (PDC), a process in which a pump photon is
annihilated to produce two entangled photons. We work in the
polarization degree of freedom and show that for any two-qubit
generation process that is both trace-preserving and
entropy-nondecreasing the concurrence $C(\rho)$ of the generated
two-qubit state $\rho$ follows an intrinsic upper bound with
$C(\rho)\leq (1+P)/2$, where $P$ is the degree of polarization of
the pump photon. We also find that for the class of two qubit
states that is restricted to have only two non-zero diagonal
elements such that the effective dimensionality of the two-qubit
state is same as the dimensionality of the pump polarization
state, the upper bound on concurrence is the degree of
polarization itself, that is, $C(\rho)\leq P$. Our work shows that
the maximum manifestation of two-particle correlations as
entanglement is dictated by one-particle correlations. The
formalism developed in this work can be extended to include
multi-particle systems and can thus have important implications
towards deducing the upper bounds on multi-particle entanglement,
for which no universally accepted measure exists.
\end{abstract}

\maketitle

The wave-particle duality, that is, the simultaneous existence of
both particle and wave properties, is the most distinguishing
feature of a quantum system. A quantum system is characterized in
terms of physical observables such as energy, momentum, etc., as
well as in terms of correlations, which, although, cannot be
measured directly like the physical observables but the degree of
which can be measured in terms of the contrast with which a system
produces interference patterns \cite{mandel&wolf, glauber1963pr,
sudarshan1963prl}. In the context of quantum systems consisting of
more than one particle, the wave-particle duality can manifest as
entanglement \cite{einstein1935pr}. Entanglement refers to
intrinsic multi-particle correlations in a system and is quite
often referred to as the quintessential feature of quantum systems
\cite{schrodinger1935}. There are many processes in which a
quantum system gets annihilated to produce a new quantum system
consisting of either equal or more number of particle. An example
is the nonlinear optical process of parametric down-conversion
(PDC), in which an input pump photon gets annihilated to produce
two entangled photons called the signal and idler photons
\cite{burnham1970prl}. Another example is the four-wave mixing
process, in which two input pump photons get annihilated to
produce two new photons \cite{boyd}. In such processes, it is
known that the physical observables get transferred in a conserved
manner \cite{burnham1970prl, mair2001nature}. For example, in
parametric down-conversion, the energy of the pump photon remains
equal to the sum of the energies of the down-converted signal and
idler photons \cite{burnham1970prl}. However, it is not very well
understood as to how the intrinsic correlations in one quantum
system get transferred to another quantum system.

One of the main difficulties in addressing questions related to
correlation transfer is the lack of a mathematical framework for
quantifying correlations in multi-dimensional systems in terms of
a single scalar quantity, although more recently there have been a
lot of research efforts with the aim of quantifying coherence
\cite{vogel2014pra,baumgratz2014prl,girolami2014prl,streltsov2015prl,yao2015pra}.
For one-particle quantum system with a two-dimensional Hilbert
space, the correlation in the system can be completely specified.
For example, polarization is a degree of freedom that provides a
two-dimensional basis and the correlations in an arbitrary state
of a one-photon system can be uniquely quantified in terms of the
degree of polarization \cite{wolf, mandel&wolf}. Two-photon
systems have a four-dimensional Hilbert space in the polarization
degree of freedom and are described by two-qubit states
\cite{kwiat1995prl}. In the last several years much effort has
gone into quantifying the entanglement of the two-qubit states
\cite{bennett1996pra, bennett1996pra2, bennett1996prl,
popescu1997pra, wootters2001qic, hill1997prl, wootters1998prl,
nielsen1999prl, torres2015prl}, and among the available
entanglement quantifiers, Wootters's concurrence
\cite{hill1997prl, wootters1998prl} is the most widely used one.
However, when the dimensionality of the Hilbert space is more than
two, there is no prescription for quantifying the correlations in
the entire system. One can at best quantify correlations in a
two-dimensional subspace \cite{born&wolf}. So, as far as
quantifying intrinsic correlations in terms of a single quantity
is concerned, it can only be done in the polarization degree of
freedom.

Different aspects of correlation transfer have previously been
investigated in degrees of freedom other than polarization
\cite{jha2008pra, jha2010pra,
jha2010prl,monken1998pra,olvera2015arxiv}. In particular,
Ref.~\cite{jha2010pra} studied correlation transfer in PDC in the
spatial degree of freedom. However, in this study, correlations
were quantified in two-dimensional subspaces only. The spatial
correlations in the pump field were quantified in terms of a
spatial two-point correlation function. For quantifying spatial
correlations of the signal and idler fields, spatial two-qubit
states with only two non-zero diagonal elements were considered.
It was then shown that the maximum achievable concurrence of
spatial two-qubit states is bounded by the degree of spatial
correlations of the pump field. In this Letter, we study
correlation transfer from one-particle to two-particle systems,
not in any restricted subspace, but in the complete space of the
polarization degree of freedom. We quantify intrinsic one-particle
correlations in terms of the degree of polarization and the
two-particle correlations in terms of concurrence.

We begin by noting that the state of a normalized
quasi-monochromatic pump field may be described by a $2\times2$
density matrix \cite{mandel&wolf} given by
\begin{equation}\label{pumpmatrix}
J \  = \ \begin{bmatrix}
\langle E_{\!_{H}}E^{*}_{\!_{H}}\rangle  &  \langle E_{\!_{H}}E^*_{\!_{V}}\rangle \\
\langle E^*_{\!_{H}}E_{\!_{V}}\rangle  &  \langle E_{\!_{V}}E^*_{\!_{V}}\rangle
\end{bmatrix},\,\,
\end{equation}
which is referred to as the `polarization matrix.' The complex random
variables $E_{\!_{H}}$ and $E_{\!_{V}}$ denote the horizontal and
vertical components of the electric field, respectively, and
$\langle\cdots\rangle$ denotes an ensemble average. By virtue of a
general property of $2\times2$ density matrices, $J$ has a decomposition of the form,
\begin{equation}\label{wolf-decomp}
 J= P\,|\psi_{\rm pol}\rangle\langle \psi_{\rm pol}| + (1-P)\,\bar{\mathds{1}},
\end{equation}
where $|\psi_{\rm pol}\rangle$ is a pure state representing a
completely polarized field, and $\bar{\mathds{1}}$ denotes the
normalized $2\times2$ identity matrix representing a completely
unpolarized field \cite{mandel&wolf}. This means that any
arbitrary field can be treated as a unique weighted mixture of a
completely polarized part and a completely unpolarized part. The
fraction $P$ corresponding to the completely polarized part is
called the degree of polarization and is a basis-invariant measure
of polarization correlations in the field. If we denote the
eigenvalues of $J$ as $\epsilon_{1}$ and $\epsilon_{2}$, then it
can be shown that $P=|\epsilon_{1}-\epsilon_{2}|$
\cite{mandel&wolf}. Furthermore, the eigenvalues are connected to
$P$ as $\epsilon_{1}=(1+P)/2$ and $\epsilon_{2}=(1-P)/2$.

\begin{figure}[t]
\includegraphics{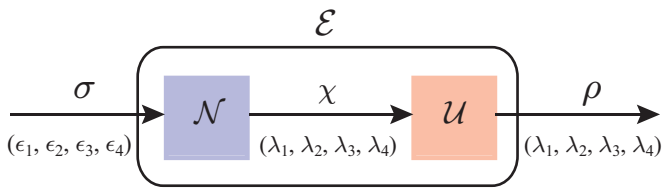}
  \caption{(color online). Modelling the generation of two-qubit states $\rho$ from $\sigma$ through a doubly stochastic process. }
  \label{fig1}
\end{figure}

We now investigate the PDC-based generation of polarization
entangled two-qubit signal-idler states $\rho$ from a
quasi-monochromatic pump field $J$ (see Fig.~\ref{fig1}). The
nonlinear optical process of PDC is a very low-efficiency process
\cite{boyd}. Most of the pump photons do not get down-converted
and just pass through the nonlinear medium. Only a very few pump
photons do get down-converted, and in our description, only these
photons constitute the ensemble containing the pump photons. We
further assume that the probabilities of the higher-order
down-conversion processes are negligibly small so that we do not
have in our description the down-converted state containing more
than two photons. With these assumptions, we represent the state
of the down-converted signal and idler photons by a $4\times 4$,
two-qubit density matrix in the polarization basis $\left\{
|H\rangle_s|H\rangle_i, |H\rangle_s|V\rangle_i,
|V\rangle_s|H\rangle_i, |V\rangle_s|V\rangle_i \right\}$. In what
follows, we will be applying some results from the theory of
majorization \cite{bhatia} in order to study the propagation of
correlations from the $2\times 2$ pump density matrix $J$ to the
$4\times4$ two-qubit density matrix $\rho$. This requires us to
equalize the dimensionalities of the pump and the two-qubit
states. We therefore represent the pump field by a $4\times 4$
matrix $\sigma$, where
\begin{align}
\sigma\equiv \left(%
\begin{array}{cc}
  1 & 0 \\
  0 & 0 \\
\end{array}%
\right)\otimes J.
\end{align}
We denote the eigenvalues of $\sigma$ in non-ascending order as
$\left(\epsilon_{1},\epsilon_{2},\epsilon_{3},\epsilon_{4}\right)\equiv\left((1+P)/2,(1-P)/2,0,0\right)$
and the eigenvalues of $\rho$ in non-ascending order as
$\left(\lambda_{1},\lambda_{2},\lambda_{3},\lambda_{4}\right)$.

Let us represent the two-qubit generation process $\sigma\to\rho$
by a completely positive map $\mathcal{E}$ (see Fig.~\ref{fig1})
such that $\rho=\mathcal{E}(\sigma)=\sum_{i}M_{i}\sigma
M^{\dagger}_{i}$, where $M_{i}$'s are the Sudarshan-Kraus
operators for the process \cite{nielsen&chuang, sudarshan1961pr,
jordan1961jmp,kraus1971}. We restrict our analysis only to maps
that satisfy the following two conditions for all $\sigma$: (i) No
part of the system can be discarded, that is, there must be no
postselection. This means that the map must be trace-preserving,
which leads to the condition that
$\sum_{i}M^{\dagger}_{i}M_{i}=\mathds{1}$; (ii) Coherence may be
lost to, but not gained from degrees of freedom external to the
system. In other words, the von Neumann entropy cannot decrease.
This condition holds if and only if the map is unital, that is,
$\sum_{i}M_{i}M^{\dagger}_{i}=\mathds{1}$. The above two
conditions together imply that the process $\sigma\to\rho$ is
doubly-stochastic \cite{nielsen2002notes}. The characteristic
implication of double-stochasticity is that the two-qubit state is
majorized by the pump state, that is $\rho\prec\sigma$. This means
that the eigenvalues of $\rho$ and $\sigma$ satisfy the following
relations:
\begin{subequations}\label{majorfull}
\begin{align}
\lambda_{1}&\leq \epsilon_{1},\label{majora}\\
\lambda_{1}+\lambda_{2}&\leq \epsilon_{1}+\epsilon_{2},\\
\lambda_{1}+\lambda_{2}+\lambda_{3}&\leq\epsilon_{1}+\epsilon_{2}+\epsilon_{3},\\
\lambda_{1}+\lambda_{2}+\lambda_{3}+\lambda_{4}&=\epsilon_{1}+\epsilon_{2}+\epsilon_{3}+\epsilon_{4}.
\end{align}
\end{subequations}
We must note that condition (i) may seem not satisfied in some of
the experimental schemes for producing polarization entangled
two-qubit states. For example, in the scheme for producing a
polarization Bell state using Type-II phase-matching
\cite{kwiat1995prl}, only one of the polarization components of
the pump photon is allowed to engage in the down-conversion
process; the other polarization component, even if present, simply
gets discarded away. Nevertheless, our formalism is valid even for
such two-qubit generation schemes. In such schemes, the state
$\sigma$ represents that part of the pump field which undergoes
the down-conversion process so that condition (i) is satisfied.

Now, for a  general realization of the process $\sigma\to\rho$,
the generated density matrix $\rho$ can be thought of as arising
from a process ${\cal N}$, that can have a non-unitary part,
followed by a unitary-only process ${\cal U}$, as depicted in
Fig.~\ref{fig1}. This means that we have $\sigma \rightarrow \chi
\equiv {\mathcal N} (\sigma) \rightarrow \rho\equiv {\mathcal
U}(\chi)$. The process $ {\mathcal N}$ generates the two-qubit
state $\chi$ with eigenvalues $\{\lambda_1, \lambda_2, \lambda_3,
\lambda_4\}$ which are different from the eigenvalues
$\{{\epsilon_1, \epsilon_2, \epsilon_3, \epsilon_4}\}$ of
$\sigma$, except when $\mathcal N$ consists of unitary-only
transformations, in which case the eigenvalues of $\chi$ remain
the same as that of $\sigma$. The unitary part ${\mathcal U}$
transforms the two-qubit state $\chi$ to the final two-qubit state
$\rho$. This action does not change the eigenvalues but can change
the concurrence of the two-qubit state. The majorization relations
of Eq.~(\ref{majorfull}) dictate how the two sets of eigenvalues
are related and thus quantify the effects due to $\mathcal N$. We
quantify the effects due to $\mathcal U$ by using the result from
Refs.~\cite{ishizaka2000pra, verstraete2001pra, wootters2001qic}
for the maximum concurrence achievable by a two-qubit state under
unitary transformations. According to this result, for a two-qubit
state $\rho$ with eigenvalues in non-ascending order denoted as
$\lambda_{1},\lambda_{2},\lambda_{3},\lambda_{4}$, the concurrence
$C(\rho)$ obeys the inequality:
\begin{align}\label{cmax}
C(\rho)\leq\mathrm{max}\{0,\lambda_{1}-\lambda_{3}-2\sqrt{\lambda_{2}\lambda_{4}}\};
\end{align}
the bound is saturable in the sense that there always exists a
unitary transformation $\mathcal U(\chi)=\rho$ for which the
equality holds true \cite{verstraete2001pra}. Now, from
Eq.~(\ref{cmax}), we clearly have $C(\rho)\leq\lambda_{1}$. And,
from the majorization relation of Eq.~(\ref{majora}), we find that
$\lambda_{1}\leq \epsilon_{1}=(1+P)/2$. Therefore, for a general
doubly-stochastic process $\mathcal E$, we arrive at the
inequality:
\begin{equation}\label{univlimit}
 C(\rho)\leq \frac{1+P}{2}.
\end{equation}
We stress that this bound is tight, in the sense that there always
exists a pair of $\mathcal N$ and $\mathcal U$ for which the
equality in the above equation holds true. In fact, the saturation
of Eq.(\ref{univlimit}) is achieved when $\mathcal N$ consists of
unitary-only process and when $\mathcal U$ is such that it yields
the maximum concurrence for $\rho$ as allowed by Eq.~(\ref{cmax}).
This can be verified, first, by noting that when $\mathcal N$ is
unitary the process $\chi=\mathcal N(\sigma)$ preserves the
eigenvalues to yield
$\left(\lambda_{1},\lambda_{2},\lambda_{3},\lambda_{4}\right)=\left((1+P)/2,(1-P)/2,0,0\right)$,
and second, by substituting these eigenvalues in Eq.(\ref{cmax})
which then yields $(1+P)/2$ as the maximum achievable concurrence.
Eq.(\ref{univlimit}) is the central result of this Letter which
clearly states that the intrinsic polarization correlations of the
pump field in PDC predetermine the maximum entanglement that can
be achieved by the generated two-qubit signal-idler states. We
note that while Eq.(\ref{univlimit}) has been derived keeping in
mind the physical context of parametric down-conversion, the
derivation does not make any specific reference to the PDC process
or to any explicit details of the two-qubit generation scheme. As
a result, Eq.(\ref{univlimit}) is also applicable to processes
other than PDC that would produce a two-qubit state from a single
source qubit state via a doubly stochastic process.

We now recall that our present work is directly motivated by
previous studies in the spatial degree of freedom for two-qubit
states with only two nonzero diagonal entries in the computational
basis \cite{jha2010pra}. Therefore, we next consider this special
class of two-qubit states in the polarization degree of freedom.
We refer to such states as `2D states' in this Letter and
represent the corresponding density matrix as $\rho^{\rm(2D)}$.
Since such states can only have two nonzero eigenvalues, the
majorization relations of Eq.(\ref{majorfull}) reduce to:
$\lambda_{1}\leq\epsilon_{1}$ and
$\lambda_{1}+\lambda_{2}=\epsilon_{1}+\epsilon_{2}=1$. Owing to
its $2\times 2$ structure, the state $\rho^{\rm (2D)}$ has a
decomposition of the form \cite{mandel&wolf},
\begin{equation}\label{wolf-decomp2}
 \rho^{\rm (2D)}=\tilde{P}|\psi^{\rm (2D)}\rangle\langle\psi^{\rm (2D)}| + (1-\tilde{P})\bar{\mathds{1}}^{\rm (2D)},
\end{equation}
where $|\psi^{\rm (2D)}\rangle$ is a pure state and
$\bar{\mathds{1}}^{\rm (2D)}$ is a normalized $2\times2$ identity
matrix. As in Eq.(\ref{wolf-decomp}), the pure state weightage
$\tilde{P}$ can be shown to be related to the eigenvalues as
$\tilde{P}=\lambda_{1}-\lambda_{2}$. It is known that the
concurrence is a convex function on the space of density matrices
\cite{wootters1998prl}, that is, $C(\sum_{i}p_{i}\rho_{i})\leq
\sum_{i}p_{i}C(\rho_{i})$, where $0\leq p_{i}\leq 1$ and
$\sum_{i}p_{i}=1$. Applying this property to
Eq.~(\ref{wolf-decomp2}) along with the fact that
$C(\bar{\mathds{1}}^{\rm (2D)})=0$, we obtain that the concurrence
$C(\rho^{\rm (2D)})$ of a 2D state satisfies $C(\rho^{\rm
(2D)})\leq \tilde{P}$. Now since
$\tilde{P}=\lambda_{1}-\lambda_{2}=2\lambda_{1}-1$, and
$\lambda_{1}\leq \epsilon_{1}$, we get $\tilde{P} \leq
2\epsilon_{1}-1=\epsilon_{1}-\epsilon_{2}=P$, or $\tilde{P}\leq
P$. We therefore arrive at the inequality,
\begin{equation}\label{2dlimit}
 C(\rho^{\rm (2D)})\leq P.
\end{equation}
Thus, for 2D states the upper bound on concurrence is the degree
of polarization itself. This particular result is in exact analogy
with the result shown previously for 2D states in the spatial
degree of freedom that the maximum achievable concurrence is
bounded by the degree of spatial correlations of the pump field
itself.\cite{jha2010pra}.

Our entire analysis leading upto Eq.~(\ref{univlimit}) and
Eq.~(\ref{2dlimit}) describes the transfer of one-particle
correlations, as quantified by $P$, to two-particle correlations
and their eventual manifestation as entanglement, as quantified by
concurrence. For 2D states, which have a restricted Hilbert space
available to them, the maximum concurrence that can get manifested
is $P$. Thus, restricting the Hilbert space appears to restrict
the degree to which pump correlations can manifest as the
entanglement of the generated two-qubit state. However, when there
are no restrictions on the available Hilbert space, the maximum
concurrence that can get manifested is $(1+P)/2$.

\begin{figure}[t]
\includegraphics{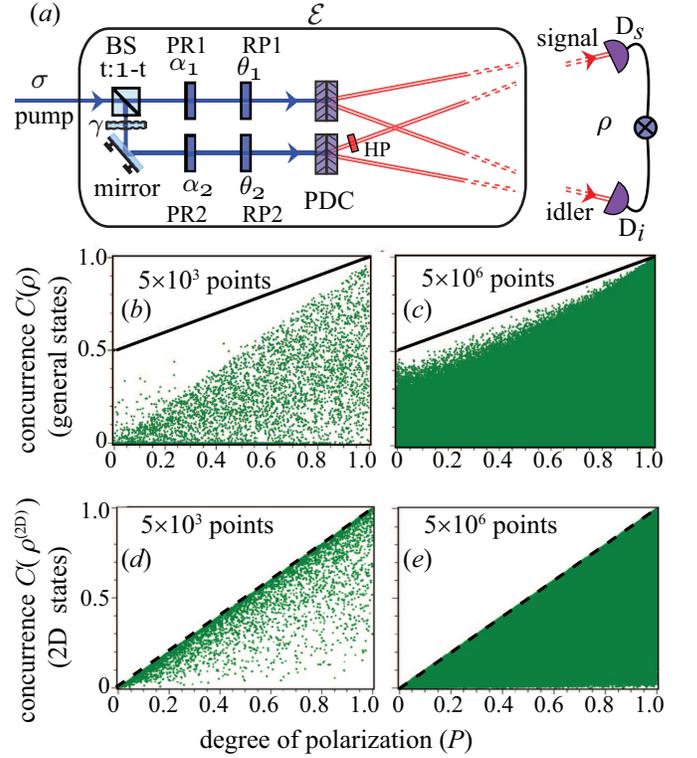}
\caption{(color online). (a) An example experimental scheme for
producing a wide range of two-qubit states. BS: beam-splitter, PR:
phase retarder, RP: rotation plate, HP: half-wave plate;
$\mathrm{D}_{s}$ and $\mathrm{D}_{i}$ are photon detectors in a
coincidence-counting setup. (b) and (c) are the scatter plots of
concurrences of states numerically generated by randomly varying
all the tunable parameters. (d) and (e) are the scatter plots of
concurrence of 2D states, numerically generated by keeping $t=1$
and varying all the remaining tunable parameters.}
\label{rank4setup2}
\end{figure}

Next, for conceptual clarity, we illustrate the bounds derived in
this Letter in an example experimental scheme shown in
Fig.~\ref{rank4setup2}(a). This scheme can produce a wide range of
two-qubit states in a doubly-stochastic manner. A pump field with
the degree of polarization $P$ is split into two arms by a
non-polarizing beam-splitter (BS) with splitting ratio $t:1-t$. We
represent the horizontal and vertical polarization components of
the field hitting the PDC crystals in arm (1) as $E_{H1}$ and
$E_{V1}$, respectively. The phase retarder (PR1) introduces a
phase difference $\alpha_{1}$ between $E_{H1}$ and $E_{V1}$. The
rotation plate (RP1) rotates the polarization vector by angle
$\theta_{1}$. The corresponding quantities in arm (2) have similar
representations. The stochastic variable $\gamma$ introduces a
decoherence between the pump fields in the two arms. Its action is
described as $\langle e^{i\gamma}\rangle=\mu\,e^{i\gamma_{0}}$,
where $\langle\cdots\rangle$ represents the ensemble average,
$\mu$ is the degree of coherence and $\gamma_{0}$ is the mean
value of $\gamma$ \cite{mandel&wolf}. The entangled photons in
each arm are produced using type-I PDC in a two-crystal geometry
\cite{kwiat1999pra}. The purpose of the half-wave plate (HP) is to
convert the two-photon state vectors $|H\rangle_{s}|H\rangle_{i}$
and $|V\rangle_{s}|V\rangle_{i}$, into
$|V\rangle_{s}|H\rangle_{i}$ and $|H\rangle_{s}|V\rangle_{i}$,
respectively. Therefore, a typical realization
$|\psi_{\gamma}\rangle$ of the two-qubit state in the ensemble
detected at $\mathrm{D}_{s}$ and $\mathrm{D}_{i}$ can be
represented as
$|\psi_{\gamma}\rangle=E_{V1}|H\rangle_{s}|H\rangle_{i}+E_{H1}|V\rangle_{s}|V\rangle_{i}
+
e^{i\gamma}\left(E_{V2}|H\rangle_{s}|V\rangle_{i}+E_{H2}|V\rangle_{s}|H\rangle_{i}\right)$.
The two-qubit density matrix is then $\rho=\langle
|\psi_{\gamma}\rangle\langle \psi_{\gamma}|\rangle=$

\footnotesize\[\!\begin{bmatrix}
\langle E_{V1}E^*_{V1}\rangle  &  \langle E_{V1}E^*_{V2}e^{-i\gamma}\rangle  &  \langle E_{V1}E^*_{H2}e^{-i\gamma}\rangle  &  \langle E_{V1}E^*_{H1}\rangle \\[4pt]
\langle E_{V2}E^*_{V1}e^{i\gamma}\rangle  &  \langle E_{V2}E^*_{V2}\rangle  &  \langle E_{V2}E^*_{H2}\rangle  &  \langle E_{V2}E^*_{H1}e^{i\gamma}\rangle \\[4pt]
\langle E_{H2}E^*_{V1}e^{i\gamma}\rangle  &  \langle E_{H2}E^*_{V2}\rangle  &  \langle E_{H2}E^*_{H2}\rangle  &  \langle E_{H2}E^*_{H1}e^{i\gamma}\rangle \\[4pt]
\langle E_{H1}E^*_{V1}\rangle  &   \langle
E_{H1}E^*_{V2}e^{-i\gamma}\rangle  &  \langle
E_{H1}E^*_{H2}e^{-i\gamma}\rangle  &  \langle
E_{H1}E^*_{H1}\rangle
\end{bmatrix}\,.\]\\[4pt]
\normalsize
For calculating the matrix elements of $\rho$, we represent the
polarization vector of the pump field before the BS as
$\left(E_{H},E_{V}\right)^{T}$ and thus write $E_{H1}$ and
$E_{V1}$ as
\begin{align}\label{fieldforms}
\begin{bmatrix}\! E_{H1} \\ E_{V1}\end{bmatrix}\!=\!\eta_{1}\!\begin{bmatrix}\!\cos{\theta_{1}} & \sin{\theta_{1}} \\
 -\sin{\theta_{1}} & \cos{\theta_{1}}\!\end{bmatrix}\!\begin{bmatrix} 1 & 0 \\
 0 & e^{i\alpha_{1}}\end{bmatrix}\!\begin{bmatrix} E_{H} \\ E_{V}\!
\end{bmatrix},
\end{align}
where $\eta_{1}=\sqrt{t}$, and the two matrices represent the
transformations by PR1 and RP1. $E_{H2}$ and $E_{V2}$ are
calculated in a similar manner, with the corresponding quantity
$\eta_{2}=\sqrt{1-t} \ e^{i\gamma}$. Without the loss of
generality, we assume $\langle E^*_{H}E_{H}\rangle=\langle
E^*_{V}E_{V}\rangle=1/2$  and $\langle E^*_{H}E_{V}\rangle=P/2$,
and calculate the matrix elements to be
\scalefont{0.9}{
\begin{multline*}
\hspace{-3.5mm}\langle E_{V1(2)}E^*_{V1(2)}\rangle=|\eta_{1(2)}|^2\big(1-P\cos\alpha_{1(2)}\sin2\theta_{1(2)}\big)/2,\\
\hspace{-10mm}\langle E_{H1(2)}E^*_{H1(2)}\rangle=|\eta_{1(2)}|^2\big(1+P\cos\alpha_{1(2)}\sin2\theta_{1(2)}\big)/2,\\
\hspace{0.2mm}\langle E_{V1(2)}E^*_{H1(2)}\rangle=|\eta_{1(2)}|^2\,P\big(\cos\alpha_{1(2)}\!\cos2\theta_{1(2)}+i\sin\alpha_{1(2)}\big)/2,\\
\hspace{-1mm}\langle
E_{V1}E^*_{V2}e^{-i\gamma}\rangle\!=\!\mu|\eta_{1}\eta_{2}|\big(\sin{\theta_{1}}\sin{\theta_{2}}+
\cos{\theta_{1}}\cos{\theta_{2}}e^{i(\alpha_{1}-\alpha_{2})}\big.\\
\big.-P\cos{\theta_{1}}\sin{\theta_{2}}e^{i\alpha_{1}}-P\sin{\theta_{1}}\cos{\theta_{2}}e^{-i\alpha_{2}}\big)e^{-i\gamma_{0}}/2,\\
\langle
E_{V1}E^*_{H2}e^{-i\gamma}\rangle\!=\!\mu|\eta_{1}\eta_{2}|\big(-\sin{\theta_{1}}\cos{\theta_{2}}+
\cos{\theta_{1}}\sin{\theta_{2}}e^{i(\alpha_{1}-\alpha_{2})}\big.\\
\big.+P\cos{\theta_{1}}\cos{\theta_{2}}e^{i\alpha_{1}}-P\sin{\theta_{1}}\sin{\theta_{2}}e^{-i\alpha_{2}}\big)e^{-i\gamma_{0}}/2,\\
\langle
E_{V2}E^*_{H1}e^{i\gamma}\rangle\!=\!\mu|\eta_{1}\eta_{2}|\big(-\cos{\theta_{1}}\sin{\theta_{2}}+
\sin{\theta_{1}}\cos{\theta_{2}}e^{-i(\alpha_{1}-\alpha_{2})}\big.\\
\big.-P\sin{\theta_{1}}\sin{\theta_{2}}e^{-i\alpha_{1}}+P\cos{\theta_{1}}\cos{\theta_{2}}e^{i\alpha_{2}}\big)e^{i\gamma_{0}}/2,\\
\hspace{-2mm}\langle
E_{H2}E^*_{H1}e^{i\gamma}\rangle\!=\!\mu|\eta_{1}\eta_{2}|\big(\cos{\theta_{1}}\cos{\theta_{2}}+
\sin{\theta_{1}}\sin{\theta_{2}}e^{-i(\alpha_{1}-\alpha_{2})}\big.\\
\big.+P\sin{\theta_{1}}\cos{\theta_{2}}e^{-i\alpha_{1}}+P\cos{\theta_{1}}\sin{\theta_{2}}e^{i\alpha_{2}}\big)e^{i\gamma_{0}}/2.
\end{multline*}}
\normalsize
Here, $t,\alpha_{1},\alpha_{2},\theta_{1},\theta_{2}$, $\mu$, and
$\gamma_0$ are the tunable parameters. We numerically vary these
parameters with a uniform random sampling and simulate a large
number of two-qubit states. Fig.\ref{rank4setup2}(b) and
Fig.~\ref{rank4setup2}(c) are the scatter plots of concurrences of
$5\times 10^3$ and $5\times 10^6$ two-qubit states, respectively,
numerically generated by varying all the tunable parameters.
Fig.\ref{rank4setup2}(d) and Fig.~\ref{rank4setup2}(e) are the
scatter plots of concurrence of $5\times 10^3$ and $5\times 10^6$
2D states, respectively, numerically generated by keeping $t=1$
and varying all the remaining tunable parameters. The solid black
lines are the general upper bound $C(\rho)= (1+P)/2$ and the
dashed black lines are the upper bound $C(\rho)= P$ for 2D states.
The unfilled gaps in the scatter plots can be filled in either by
sampling more data points or by adopting a different sampling
strategy. To this end, we note that one possible setting for which
the general upper bound can be achieved is:
$t=0.5,\theta_{1}=-\pi/4,\theta_{2}=0,\alpha_{1}=\pi/2,\alpha_{2}=\pi,
\mu=1$ and $\gamma_{0}=0$.

In conclusion, we have investigated how one-particle correlations
transfer to manifest as two-particle correlations in the physical
context of PDC. We have shown that if the generation process is
trace-preserving and entropy-nondecreasing, the concurrence
$C(\rho)$ of the generated two-qubit state $\rho$ follows an
intrinsic upper bound with $C(\rho)\leq (1+P)/2$, where $P$ is the
degree of polarization of the pump photon. For the special class
of two-qubit states $\rho^{\rm (2D)}$ that is restricted to have
only two nonzero diagonal elements, the upper bound on concurrence
is the degree of polarization itself, that is, $C(\rho^{\rm
(2D)})\leq P$. The surplus of $(1+P)/2-P=(1-P)/2$ in the maximum
achievable concurrence for arbitrary two-qubit states can be
attributed to the availability of the entire $4\times 4$
computational space, as opposed to 2D states which only have a
$2\times 2$ computational block available to them. We believe
these results can have two important implications. The first one
is to understand from a fundamental perspective, whether or not
correlations too follow a quantifiable conservation principle just
as physical observables such as energy, momentum do. The second
one is that this formalism provides a systematic method of
deducing an upper bound on the correlations in a generated quantum
system, purely from the knowledge of the correlations of the
source. In the light of the recent experiment on generation of
three-photon entangled states from a single source photon
\cite{hamel2014natphoto}, this formalism may prove useful in
determining upper bounds on the entanglement of such multipartite
systems, for which no well-accepted measure exists. This
alternative approach based on intrinsic source correlations could
complement the existing information-theoretic approaches
\cite{bennett1996pra, bennett1996pra2, bennett1996prl,
popescu1997pra, wootters2001qic, hill1997prl, wootters1998prl,
nielsen1999prl, torres2015prl} towards quantifying entanglement.

GK acknowledges helpful discussions on the Physics StackExchange
online forum. AKJ acknowledges financial support through an
initiation grant from IIT Kanpur.

\end{document}